# Plasmonless polarization-selective metallic Semi-coaxial Aperture Arrays in the visible range

**Abdoulaye Ndao*[1, 2], Roland Salut[1] and Fadi I. Baida[1]**

*1- Institut FEMTO-ST, UMR CNRS 6174, Université de Franche-Comté, Département d'Optique P.M. Duffieux, 15B avenue des Montboucons, 25030 Besançon cedex, France*
*2- Department of Electrical and Computer Engineering, University of California San Diego, La Jolla, California 92093-0407, USA*
abndao@eng.ucsd.edu

**Abstract**

In this letter, we perform numerical and experimental studies of the optical response of an original configuration based on enhanced transmission through guided mode based metamaterials. The proposed structure is inspired by annular aperture array (AAA) where the cylindrical symmetry is broken in order to acquire a polarization-sensitive metasurfaces. The experimental results, which are in good agreement with numerical simulations, demonstrate that the structure acts as a polarizer exhibiting an extinction ratio of (15:1) with a maximum transmission coefficient up to 85% which is more efficient than what it is expected with a typical plasmonic resonance.

Since the discovery of the electromagnetic surface plasmon (SP) wave by Kretschmann [1], all the experiments proving the existence of this excitation were based on the optimization of the absorption of the incident electromagnetic field by the metallic structure. In fact, when SP is efficiently excited, it pumps all the incident energy so that both transmission and reflection are zero. At this time, the surface wave is excited and the light propagates along the metal-dielectric interface until all the energy is dissipated due to the metal absorption. This is in direct contradiction with the Ebbesen [2] experiment where extraordinary transmission, and not absorption, is associated with the excitation of the SP. In fact, the metallic structure (aperture array) of Ebbesen plays the role of an optical tunnel for the surface plasmon wave. The extraordinary transmission can perhaps be associated with the excitation of a SP but this excitation is not very efficient since a part of the incident power is not converted to a plasmonic wave. Hence, we name this effect "frustrated surface plasmon resonance".

Research groups around the world have proposed to investigate the large potential of metamaterials [3,4], but the most proposed solutions are based on surface plasmon resonance (SPR) within metallic nanostructures [5, 6]. Nevertheless, a more interesting solution consists of the excitation of specific eigenmodes of the structure exhibiting less absorption in comparison to the SPR. In this context, AAA (Annular Aperture Arrays) was first proposed [7] to compete with the Ebbesen structures demonstrating larger transmission coefficients (at least 4 times) due to the propagation of guided modes inside the apertures. Optical properties of these modes were extensively studied both theoretically and experimentally for a wide range of applications such as based color filters for the visible range [8], enhancement of second harmonic generation [9], beam angular filtering [10] and broadband enhanced transmission metamaterials for visible [11-14] and infrared applications [15]. The main feature discrepancy between guided mode-based and surface plasmon resonance-based extraordinary transmission is the spectral bandwidth, a property that will be exploited here to propose metamaterial acting as a spectral broadband polarizer in the visible to near infrared (NIR) range. Note that other geometries of nano-waveguide (ellipses [16] or rectangles [17]) exist and were proposed in the context of polarization control of the transmitted beam. Nevertheless, the SPR-guided mode coupling can negatively affect the transmission amplitude and modify its spectral bandwidth. Moreover, the spectral position ($\lambda_A$) of these anomalies (Rayleigh and Wood) is governed by the following equation:

$$\lambda_A^2 \left(\frac{m^2}{p_x^2} + \frac{l^2}{p_y^2}\right) + 2\sqrt{\varepsilon_1}\lambda_A sin\theta \left(\frac{m cos\phi}{p_x} + \frac{l sin\phi}{p_y}\right) + (\varepsilon_1 \sin^2 \theta - \alpha^2(\lambda_A)) = 0 \quad (1)$$

where $\varepsilon_1$ and $\varepsilon_2$ are the relative dielectric permittivities of the incidence (substrate) and transmission (superstrate) media respectively, *px* and *py* are the periods along the *Ox* and *Oy* directions and *m* and *l* are integers of the diffraction order in the *Ox* and *Oy* directions. ϕ and θ are Euler angles that define the incidence direction. α is a parameter that can take four different values:

$\alpha = n_1 = \sqrt{\varepsilon_1}$   Rayleigh anomalies at substrate metal interface

$\alpha = \sqrt{\frac{\varepsilon_m \varepsilon_1}{\varepsilon_m + \varepsilon_1}}$   Wood anomalies at substrate-metal interface

$\alpha = n_2 = \sqrt{\varepsilon_1}$   Rayleigh anomalies at substrate metal interface

$\alpha = \sqrt{\frac{\varepsilon_m \varepsilon_2}{\varepsilon_m + \varepsilon_2}}$   Wood anomalies at substrate-metal interface

We will limit our study to normal incident beams linearly polarized along a single direction (ϕ = θ = $0^0$) and we consider a grating with the same period *px* = *py* = *p* so that equation 1 becomes:

$$\lambda_A^2(m^2 + l^2) - \alpha^2 p^2 = 0$$

According to equation 2, the (*m* = ±1; *l* = 0) and (*m* = 0; *l* = ±1) are the first four diffracted orders that can disturb (or couple) the excitation of the fundamental guided mode inside the apertures. In the case of a metallic layer with small thickness (*h* < 150*nm*), the excitation of the guided mode leads to an efficient transmission over a wide spectral range (broader transmission peak) around its cutoff wavelength [18] and can then overlap the SPR. Thus, their coupling leads to the emergence of a Fano-like shape in the transmission spectrum. Unfortunately, a transmission dip occurs at a spectral position that greatly depends on the periodicity of the structure associated with the SPR excitation (see eq. 1). On the contrary, the transmission properties of the guided mode are quasi-independent of the periodicity especially if the metal thickness is small [19] resulting in a transmission peak at the vicinity of the cutoff spectral position of the mode. In the case of an infinitely long coaxial waveguide, the latter can be written as [20]:

$$\lambda_c = \pi(R_o + R_i) + a$$

Where *Ro* and *Ri* are the outer and inner radii of the aperture respectively and the term α is a Reds-hift term induced by the metal absorption. The value of α depends on the metal dispersion and on the structure geometry (especially when the gap *Ro* −*Ri* is smaller than 200 *nm* [20]). It is also easy to understand that for small period values, a coupling between two adjacent apertures can occur across the metal thickness due to the light penetration inside the metal. This effect will be discussed below.

To illustrate the coupling properties between the vertical guided mode and the horizontal surface plasmon mode, we have designed the structure depicted on figure 1 and named Semi- Coaxial Aperture Array (SCAA). The latter (1) draws on the AAA structure mentioned above where the degeneracy of the $TE_{11}$ mode, due to the cylindrical symmetry of the apertures, is deliberately broken by leaving a small metallic part along one aperture diameter along a periodic direction (here the *Ox*−direction). This structure was proposed and numerically studied in ref. [21] highlighting out its polarization sensitivity. In the context of the present study, this polarization sensitivity is experimentally demonstrated and is exploited to select two distinct operation regimes by simply rotating the polarization direction of the incident wave: A first one where only the SPR is excited and a second regime where Fano resonance occurs due to interference between the horizontal SPR mode and the in-aperture vertical guided mode.

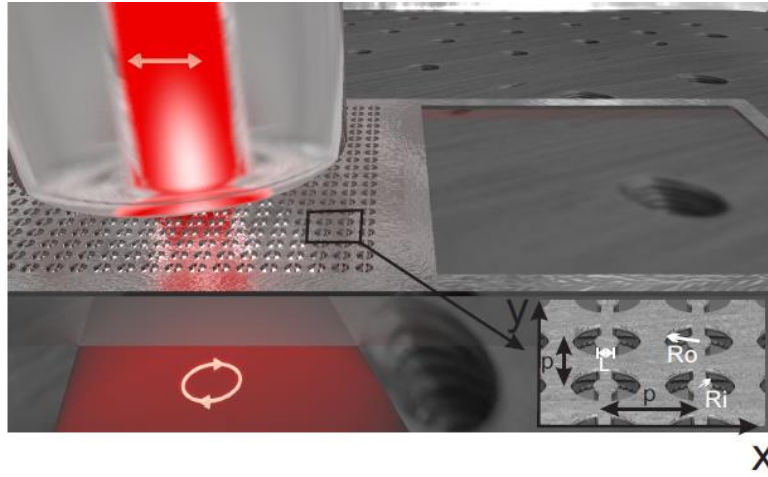

Figure 1 Schematic of the semi-coaxial aperture array (SCAA) engraved into metallic (silver) layer deposited on glass substrate. The reference zone has the same dimension as the whole array. Inset on the top right indicates the variables associated with each geometrical parameter of an aperture

Geometrical parameters of this structure are then optimized in order to get spectral broadband transmission coefficient in the visible-NIR range. To this end, a home-made 3D-FDTD (Finite Difference Time Domain) code is utilized to simulate the optical response of our structure. It integrates an efficient analytical model allowing an accurate description of the metal dispersion properties (silver for instance), namely the Drude-critical points model [22, 23] that is adapted to the experimental data of ref. [24].

*1.1. Numerical study*

The simulations allow the determination of the most appropriate geometrical parameters (thickness, diameters and metal nature) to experimentally demonstrate an enhanced transmission assisted by a $TE_{11}$ guided mode in view of obtaining an efficient nano-polarizer. The design of the structure imposes two conditions: the first one is to affix the air gap (the difference between inner and outer radii) to be sufficient in order to facilitate the technological challenge of fabrication while the second is to choose the value of the period to allow Woods and/or Rayleigh's anomalies to be coupled with the guided mode. In fact, at normal incidence and according to Eq. 2, the anomaly exhibiting the largest value of wavelength corresponds to the SPR induced by the diffracted orders ($m = 1; l = 0$) or ($m = 0; l = 1$). Thus, we first fixed $h = 100$ *nm*, $L = 50$ *nm*, $Ro = 150$ *nm* and $Ri = 50$ *nm* and have varied the period value between $p = 320$ *nm* and $p = 900$ *nm* in order to see the evolution of the coupling between the fundamental guided mode ($TE_{11}$) and this SPR mode. Nevertheless, the guided mode is only excited in $x-$polarization (figure 2a) while the SPR exists also for the $y-$polarization incident electric field (figure 2b). The cutoff wavelength of the $TE_{11}$ mode was calculated through an order-N spectral FDTD algorithm [25] incorporating the same dispersion model of Drude critical points. The obtained value is $\lambda c = 756$ *nm* leading to $\alpha = 186.48$ *nm* in equation 3. As expected, we can see from figure 2 that the transmission is more important in the case of the guide mode excitation and reaches 92% while the maximum for the $y-$polarization is of 54%. This latter value, that occurs for small period values, is somewhat surprising (too large) but is due to a hybrid guided mode which is different from the pure $TE_{11}$ mode resulting from the coupling between the apertures themselves across the metal thickness. In fact for $p < 350$ *nm* the distance between two consecutive apertures becomes of the same order as the magnitude of twice the metal skin depth ($\approx 2\times 25$ *nm*). The white dots and circles plotted on figures 2 indicate the SPR spectral position as calculated from equation 2. At first glance, they seem to be superimposed on the transmission dip positions for both $x-$ and $y-$ polarizations meaning that the coupling effect is very small. However, by determining the exact position of the transmission dip ($\lambda d$) we calculate the difference between them and the theoretical value given by equation 2. Let $\Delta = \lambda A - \lambda d$ be this difference.

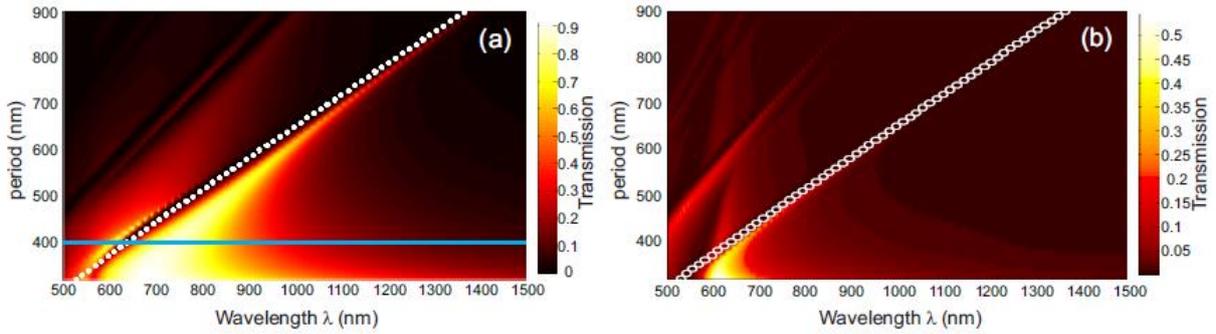

Figure 2 Transmission through SCAA calculated using a custom FDTD code as a function of both wavelength and period. Comparison between simulation and the analytical model given by equation 2 for a linearly polarized plane wave along the x –directions is presented in (a) while the y-polarization is given in (b). The other geometrical parameters are Ri=50 nm, Ro=150 nm, P=400 nm and L=50 nm.

Values of $\Delta$ are depicted on figure 3a as a function of the wavelength (bottom horizontal axis) and as the period value (top axis) for $x-$ polarized incident plane wave when both guided and SPR are excited (red line and stars) and in $y-$ polarization when only the SPR exists (blue line and stars). We can clearly show that the signature of the coupling effect is more pronounced when the $TE_{11}$ mode is excited. Nevertheless, the two curves converge toward $\Delta \simeq 0$ nm when $p$ increases. This clearly demonstrates that the excitation of the SPR spectral position corresponds to a transmission dip and not to a maximum transmission. Nonetheless, the presence of the SPR is necessary to induce a fairly little important transmission (to the right of the dip) induced by a Fano resonance resulting from the coupling between the SPR mode and residual transmission corresponding to a guided mode that is spectrally located far from the SPR. Figure 3b presents the transmission spectrum for $p = 400$ nm (blue cross section line made over the map of figure 2a). This value of the period was chosen to be fabricated because it allows efficient transmission over a wide spectral range including the visible domain. With this value, Eq. 2 leads to $\lambda SPR(l = 1; m = 0) = 638.5$ nm while the dip position in the transmission spectrum is located at $\lambda d = 635.63$ nm. This small discrepancy ($\Delta = 2.87$ nm) is due to the presence of the guided mode that induces a blue shift of the SPR. The value of $L = 50$ nm used in the previous study was determined to be the minimum value that insures a good extinction ratio (ER) between the two orthogonal polarization directions.

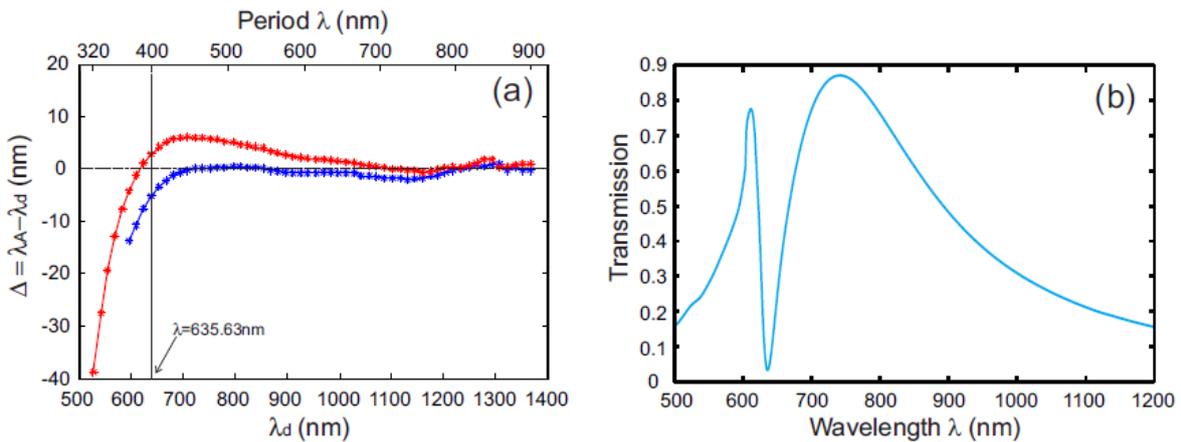

Figure 3 (a) Difference between the expected SPR spectral position ($\lambda A$) from equation 2 and the transmission dip (ld) as function of the dip position (bottom horizontal axis ld ) and as the SCAA period(top axis) for x-polarization (red curve) and y-polarization (blue curve). (b) Transmission spectrum as a function of the wavelength in the case of a period value p=400 nm. The others geometrical parameters in (b) are Ri=50 nm, Ro=150 nm, L=50 nm.

Figure 4 shows the obtained numerical spectra when $L$ varies from 10 nm to 70 nm for the two polarization states (along $Ox$ and $Oy$ directions). We can clearly see that the $TE_{11}$ guide mode is still efficiently excited for the $x$-polarization while smaller transmission occurs for the $y$-polarization. As mentioned previously, at the SPR wavelength, the transmission coefficient is zero for both the two polarization states. One time again, this denotes the negative role of the SPR in the transmission process [26]. On the one hand, the spectral position of the SPR is quite independent of the value of $L$. The transmission dip position only varies from 632.95 $nm$ to 629.45 $nm$ for $x$-polarization and from 640.45 to 639.92 $nm$ for the $y$-polarization. On the other hand, the transmission peak position appears to be more sensitive to $L$ because its position varies from 724 $nm$ to 746 $nm$ when $L$ varies from 30 $nm$ to 70 $nm$. Fortunately, its quality factor is very weak and leads to a very small relative variation of the transmitted energy. According to Figure 4, and taking into account the fabrication constraints, i.e. Focused Ion Beam (FIB) resolution, we estimated $L = 50$ $nm$ is sufficient to design efficient nano-polarizer with an average ER of 16 for $\lambda \in [700-1600]$ $nm$ (see inset of figure 4). Specifically, transmission coefficients of $Tx = 0.75$ and $Ty = 0.043$ are obtained at an operation wavelength of $\lambda = 794$ $nm$ leading to an ER of 17.44. Thereby, the proposed structure almost behaves as a linear polarizer in that spectral range. More importantly, it allows distinguishing the case of a specific SPR excitation (polarization along the $y$-direction) of the case where the SPR is coupled with the guided vertical mode (incident polarization directed toward the $x$-direction).

Nevertheless, until now the metal thickness was arbitrarily fixed to $h = 100$ $nm$ which corresponds to a common value recently used in several experimental configurations [11, 27, 28]. In fact, the transmission properties of such metamaterial are independent of this parameter for SPR or guided mode excitation. This is clearly demonstrated through the calculated spectra presented in figure 5 where the parameter $h$ was varied from $h = 80$ $nm$ to $h = 170$ $nm$. For this range of the thickness, the transmission dip position (SPR excitation) only varies from 631 $nm$ to 633 $nm$. At the same time, the transmission efficiency at the $TE_{11}$ guided mode excitation decreases from 0.8875 to 0.798. However, the quality factor of the transmission peak slightly increases passing from 2.41 to 3.88 when $h$ grows from 80 $nm$ to 170 $nm$. Nonetheless, the fabrication constraints (maximum aspect ratio of 1 to minimize the conical effect [11] of the FIB metal milling) impose an $h$ value around 100 $nm$. In addition, the smaller $h$ is, the smaller the metal absorption. All the geometrical parameter of the structure are now fixed ($Ri = 50$ $nm$, $Ro = 150$ $nm$, $L = 50$ $nm$ and $p = 400$ $nm$.). In the following, we will discuss its fabrication and its characterization.

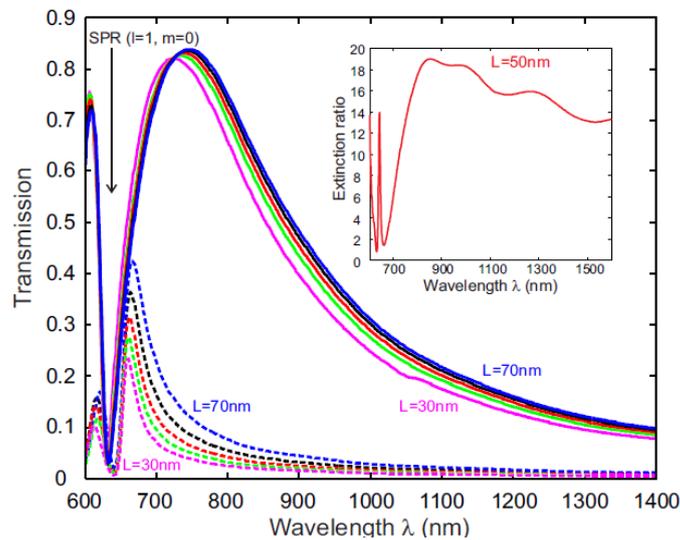

Figure 4 Theoretical zero-order transmission spectra through an SCAA for different values of L for both x-(solid lines) and y-(dashed lines) polarized incident beam impinging the structure from the glass substrate at normal incidence with Ri=50 nm, Ro=150 nm, h=100 nm and p=400 nm. The inset gives the extinction ratio(ER) for L=50 nm as a function of the illumination wavelength.

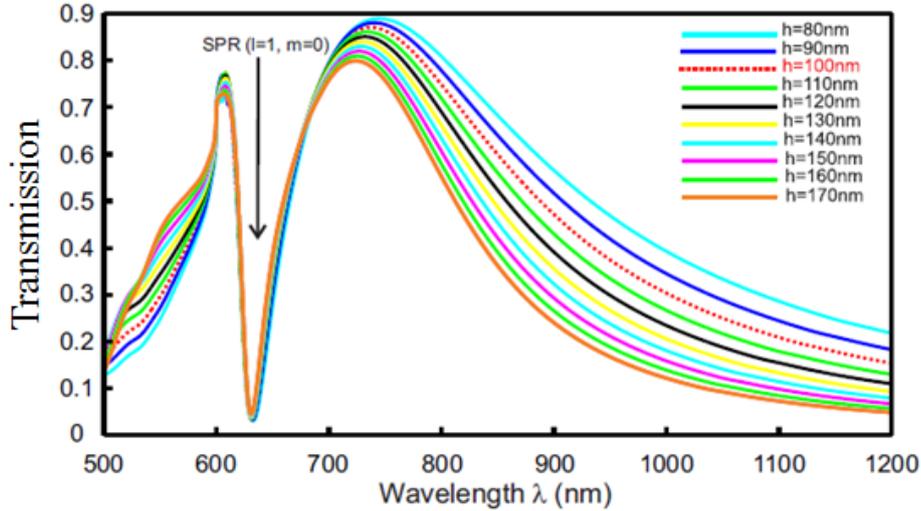

Figure 5 Zero-order theoretical transmission spectra through a Semi coaxial structure versus the thickness h of the silver for a transverse magnetic (X) polarization illuminated at normal incidence. The other geometrical parameters are Ri=50 nm, Ro=150 nm, L= 50 nm and p= 400 nm.

### *1.2. Fabrication*

FIB milling combined with a very accurate metal deposition by evaporation is used to realize the fabrication of the SCAA. First, a 5 *nm* thin chromium layer is deposit on the glass substrate (refractive index $n = 1.5$) as an adhesion layer. Next, a silver film (thickness $h = 100$ *nm*) is deposited by evaporation. Finally, the apertures are obtained by FIB milling of the metallic layer. A SEM image of the fabricated sample is shown in fig.6. The Whole matrix is $8\times 8\mu m^2$ composed of $20\times 20$ apertures. Despite a few minor imperfections (see small yellow circles that indicate un-etched apertures), the SEM images show well defined apertures and periodicity along both *x*- and *y*- directions.

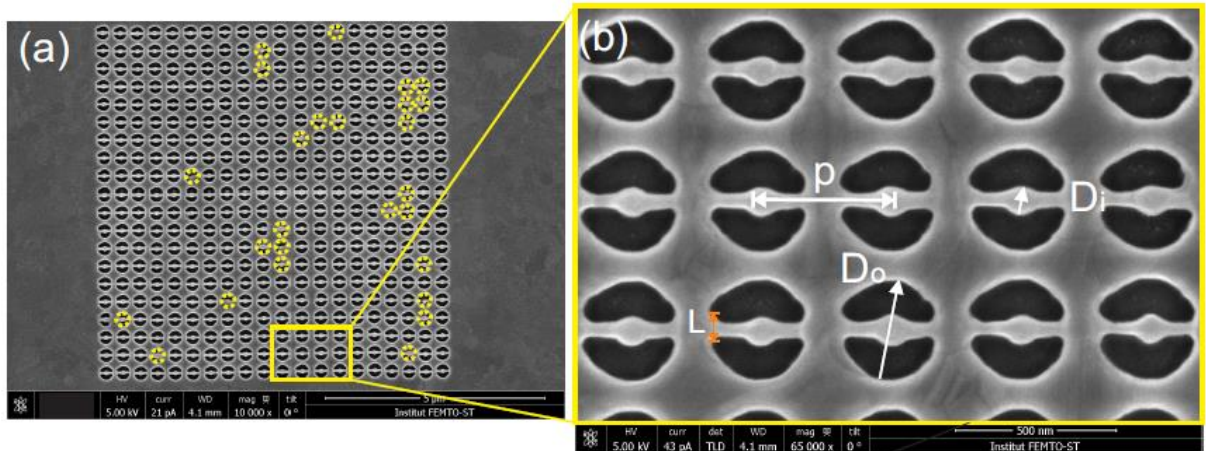

Figure 6 (a) SEM top view image of the studied AAA (20×20 apertures) engraved in silver film. The yellow circles indicate some of the apertures that are not completely etched by FIB milling. (b) Zoom-in over 5x3 patterns.

### *1.3. Characterization*

To record the transmission spectra, the sample is illuminated at normal incidence by a supercontinuum white light source (Leukos-SM OEM, from 320 *nm* to 2400 *nm*) after it is collimated and linearly polarized (*x* or *y*- direction). The zero order transmission is then propagated to a spectrometer through a cleaved multimode fiber (core diameter 62.5μm) placed above the SCAA matrix and connected to an optical spectrometer (USB 2000 by Ocean Optics). We recall here that the transmission coefficients are theoretically and experimentally defined as the ratio between the diffracted zero order transmission through the SCAA to a reference area consisting in a square aperture having the same lateral size as the whole SCAA and engraved in the same metallic film. The transmission spectra for different angles of polarization (from ϕ = 0° to ϕ = 90°) are depicted in fig.7. To experimentally validate the theoretical predictions and the protocol fabrication, we compared in fig.7 the theoretical and experimental results for both polarization (Einc //Ox and Einc // Oy). In the case of the Ox polarization, taking into account the measured values of the radii ($R_e$ = 141 nm, R = 50 nm) and the asymmetry of the L value (L = 50 nm on one side and 36.2 nm of each other), we observe a good agreement between theory and experiment in terms of position and intensity (about 80% transmission of the guided mode $TE_{11}$). We also see good agreement of peak shape with a slight difference of about 5 nm position. Certainly, fabricated structures do not present any serious defects except the asymmetrical shape of the metal portion that defines the semi-coaxial aperture. Nevertheless, one notices the slope change in the experimental spectra that occurs due to the clipping of the optical spectrometer at higher wavelength values (λ >= 800 nm). As shown in the study according to L on figure 4, small variations of this parameter affects very weakly on the amplitude and the position of the transmission peak. This proves the robustness of the design of the SCAA. In contrast to x−polarization, the transmission obtained is very low (of about 15%) for the y−polarized incident beam. In addition, in terms of position, one sees a much larger gap between the theoretical and the experimental spectra due to the experimental conditions as the polarization changes (estimated error to 2 degrees). These results show that this structure allows us to control the confinement selectively depending on the polarization and thereby obtain a transmission two times higher than that associated with rectangular slots [17].

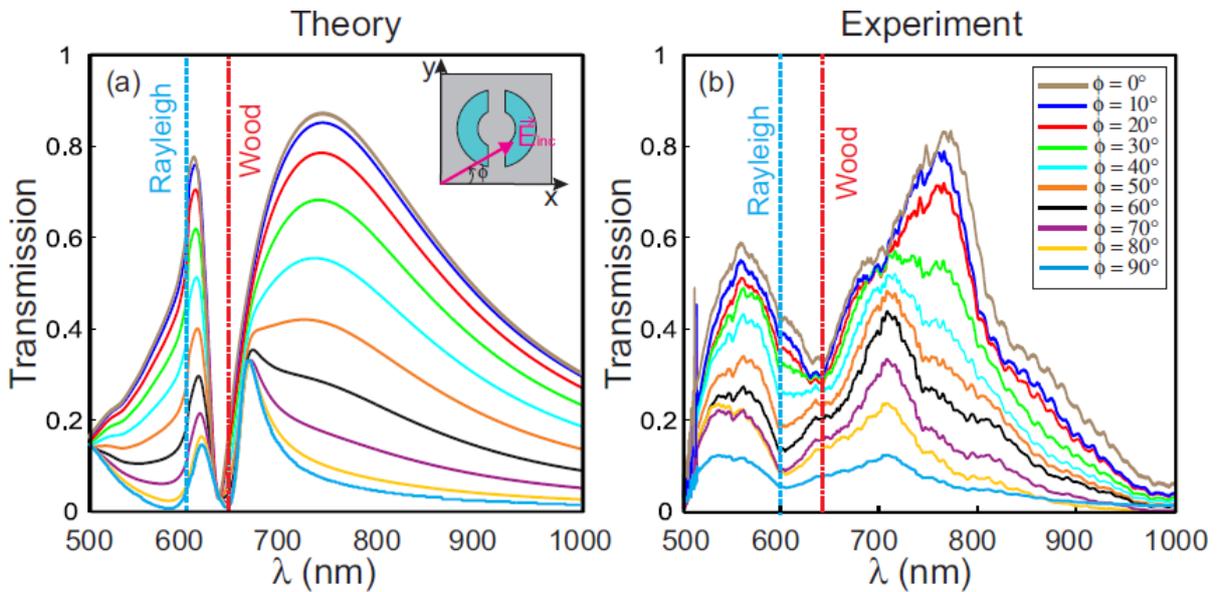

Figure 7 Zero-order (a) theoretical and (b) experimental transmission spectra of the semi-coaxial annular apertures structure as a function of the polarization: Note that the simulations are based on a home-made 3D FDTD code.

Finally, we plot on figure 8 the experimental ER calculated as $T(\phi = 0°)/T(\phi = \pi/2)$. The obtained result agree well with the theoretical one depicted as inset in figure 4 and corresponding to the same geometry (L = 50 nm for instance). Unfortunately, the maximum value of the ER is smaller than the theoretical one due to the fact that, as mentioned previously, some apertures (exactly 44) were not completely etched and did not participate to the transmission coefficient for ϕ = 0°. However, an ER of 15 is obtained at λ = 750 nm that corresponds to the position of the guided mode-based transmission peak.

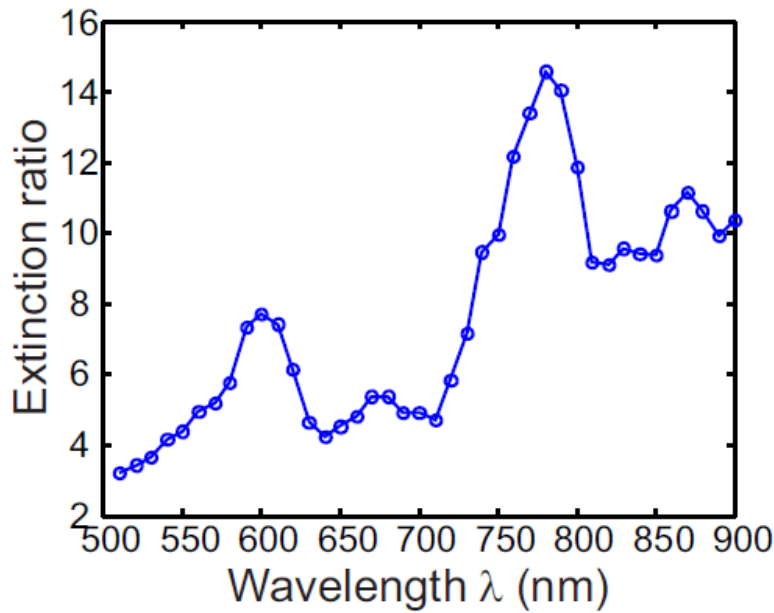

Figure 8 Experimental ER calculated over a restricted spectral range corresponding to the OSA bandwidth

**2. Conclusion**

We numerically and experimentally demonstrated high polarization-sensitive metamaterial with very thin metallic layer based on enhanced transmission through guided mode excitation inside the apertures. The experimental results, which are in good agreement with numerical simulations, demonstrate that the structure acts as a polarizer exhibiting an extinction ratio of (15:1) with a maximum transmission coefficient of 85% which is more efficient than what can be expected with plasmonic resonance. This nano-polarizer function is a promising solution for nearfield polarization microscopy (selective detection of the transverse components of the electric near-field). It is then sufficient to integrate one semi-coaxial aperture at the end of a Scanning Near-field Optical Microscope (SNOM) metal coated probe to get much better efficiency than that of a structure with an elliptical cross-section as suggested by Gordon et al. [16].

**Acknowledgments**

We acknowledge financial support by the Labex ACTION program (Contract No. ANR-11- LABX-0001-01). We are grateful to B. Guichardoz and M. Suarez for help in the preparation of the figures. Computations have been performed on the supercomputer facilities of the "Mésocentre de calcul de Franche-Comté"

**References**

[1] Kretschmann and H. Raether, "Radiative Decay of the Non-Radiative Surface Plasmons Excited by Light," Z. Naturforsch 239, 2135 (1963).
[2] T. Ebbesen, H. Lezec, H. Ghaemi, T. Thio, and P. Wolff, "Extraordinary optical transmission through sub-wavelength hole arrays," Nature 391, 667–669 (1998).
[3] H. Caglayan, I. Bulu, M. Loncar, and E. Ozbay, "Experimental observation of subwavelength localization using metamaterial-based cavities," Opt. Lett. 34 (2009).
[4] Z. Liu, H. Lee, Y. Xiong, C. Sun, and X. Zhang, "Far-Field Optical Hyperlens Magnifying Sub-Diffraction-limited Objects," Science 315, 1686 (2007).
[5] M. Sukharev and T. Seideman, "Phase and polarization control as a route to plasmonic nanodevices," Nano Lett. 6(4), 715–719 (2006).
[6] M. Aeschlimann, M. Bauer, D. Bayer, T. Brixner, F. J. G. de Abajo, W. Pfeiffer, M. Rohmer, C. Spindler, and F. Steeb, "Adaptive subwavelength control of nano-optical fields," Nature 446, 301 (2007).
[7] F. Baida and D. Van Labeke, "Light transmission by subwavelength annular aperture arrays in metallic films," Opt. Commun. 209, 17–22 (2002).
[8] G. Si, Y. Zhao, H. Liu, S. Teo, and M. Zhang, "Annular aperture array based color filter," Appl. Phys. Lett. 99, 033,105 (2011).

[9] E. Barakat, M.-P. Bernal, and F. I. Baida, "Doubly resonant Ag–LiNbO3 embedded coaxial nanostructure for high second-order nonlinear conversion," J. Opt. Soc. Am. B 30(7), 1975–1980 (2013).
[10] A. Roberts, "Beam transmission through hole arrays." Opt. Express 18, 2528 (2010).
[11] A. Ndao, A. B. B. Salut, and F. I. Baida, "Slanted annular aperture arrays as enhanced-transmission metamaterials: Excitation of the plasmonic transverse electromagnetic guided mode," Appl. Phys. Lett. 103, 211,901(2013).
[12] A. Ndao, J. Salvi, R. Salut, M.-P. Bernal, T. Alaridhee, A. Belkhir, and F. I. Baida, "Resonant optical transmission through sub-wavelength annular apertures caused by a plasmonic transverse electromagnetic (TEM) mode," J. Opt. 16(12), 125,009 (2014).
[13] T. Alaridhee, A Ndao, M.-P Bernal, Evgueni Popov, Anne-Laure Fehrembach, F. I. Baida "Transmission properties of slanted annular aperture arrays. Giant energy deviation over sub-wavelength distance", Opt. Express, (23), 11687-11701 (2015).
[14] M. Hamidi, C. Chemrouk, A. Belkhir, Z. Kebci, A. Ndao, O. Lamrous, F.I. Baida "SFM-FDTD analysis of triangular-lattice AAA structure: Parametric study of the TEM mode," Optics Communications 318 47–52 (2014).
[15] W. Fan, S. Zhang, K. J. Malloy, and S. R. J. Brueck, "Enhanced mid-infrared transmission through nanoscale metallic coaxial-aperture arrays," Opt. Express 13(12), 4406–4413 (2005).
[16] R. Gordon, A. G. Brolo, A. McKinnon, A. Rajora, B. Leathem, and K. L. Kavanagh, "Strong Polarization in the Optical Transmission Through Elliptical Nanohole Arrays," Phys. Rev. Lett. 92(3), 37,401 (2004).
[17] A. Ndao, Q. Vagne, J. Salvi, and F. I. Baida, "Polarization sensitive sub-wavelength metallic structures: Toward near-field light confinement control," Appl. Phys. B 106(4), 857–862 (2012).
[18] F. I. Baida, D. Van Labeke, G. Granet, A. Moreau, and A. Belkhir., "Origin of the super-enhanced light transmission
through a 2-D metallic annular aperture array: a study of photonic bands," Appl. Phys. B 79(1), 1–8 (2004).
[19] F. I. Baida and D. Van Labeke, "Three–Dimensional Structures for Enhanced Transmission Through a Metallic Film: Annular Aperture Arrays," Phys. Rev. B 67, 155,314 (2003).
[20] F. I. Baida, A. Belkhir, D. Van Labeke, and O. Lamrous, "Subwavelength metallic coaxial waveguides in the optical range: Role of the plasmonic modes," Phys. Rev. B 74(20), 205,419 (2006).
[21] F. Baida, Y. Poujet, B. Guizal, and D. Van Labeke, "New design for enhanced transmission and polarization control through near-field optical microscopy probes," Optics Commun. 256, 190–195 (2005).
[22] A. Vial, T. L. L. Dridi, and L. L. Cunff, "A new model of dispersion for metals leading to a more accurate modeling of plasmonic structures using the FDTD method," Appl. Phys. A 103(3), 849–853 (2011).
[23] M. Hamidi, F. I. Baida, A. Belkhir, and O. Lamrous, "Implementation of the critical points model in a SFMFDTD code working in oblique incidence," J. Phys. D: Appl. Phys. 44(24), 245,101 (2011).
[24] E. Palik, *Handbook of Optical Constants of Solids. Acad. Press.,* (Acad. Press., 1985).
[25] C. Chan, Q. Yu, and K. Ho, "Order-N Spectral Method for ElectromagneticWaves," Phys. Rev. B 51(23), 16,635– 16,642 (1995).
[26] Q. Cao and P. Lalanne, "Negative role of surface plasmon in the transmission of metallic gratings with very narrow slits," Phys. Rev. Lett. 88(5), 057,403–1 (2002).
[27] Y. Poujet, J. Salvi, and F. I. Baida, "90% Extraordinary optical transmission in the visible range through annularaperture metallic arrays," Opt. Lett. 32, 2942–2944 (2007).
[28] E. Barakat, M.-P. Bernal, and F. I. Baida, "Theoretical analysis of enhanced nonlinear conversion from metallodielectric nano-structures," Opt. Express 20, 16,258–16,268 (2012).